\documentclass[prl,showpacs,amsmath,amssymb,twocolumn,10pt]{revtex4}
\usepackage{enumerate}
\usepackage{color}
\usepackage{amsthm}
\usepackage{dcolumn}
\usepackage{bm}
\usepackage{graphicx}

\begin{document}

\newtheorem*{corollary}{Corollary}
\newtheorem{definition}{Definition}
\newtheorem{example}{Example}
\newtheorem{lemma}{Lemma}
\newtheorem{proposition}{Proposition}
\newtheorem{theorem}{Theorem}
\newtheorem*{fact}{Fact}
\newtheorem{property}{Property}
\newcommand{\bra}[1]{\langle #1|}
\newcommand{\ket}[1]{|#1\rangle}
\newcommand{\braket}[3]{\langle #1|#2|#3\rangle}
\newcommand{\ip}[2]{\langle #1|#2\rangle}
\newcommand{\op}[2]{|#1\rangle \langle #2|}
\newcommand{\Sl}{\overrightarrow{S}_{LOCC}}
\newcommand{\Ss}{\overrightarrow{S}_{Sep}}
\newcommand{\Slp}{\overrightarrow{S}^{prod}_{LOCC}}
\newcommand{\Ssp}{\overrightarrow{S}^{prod}_{Sep}}

\newcommand{\tr}{{\rm tr}}
\newcommand{\supp}{{\it supp}}

\newcommand{\slocc}{\overset{\underset{\mathrm{SLOCC}}{}}{\longrightarrow}}
\newcommand{\comments}[1]{}
\newcommand{\W}{|W\rangle}
\newcommand{\WW}{\W^{\otimes2}}
\newcommand{\rk}{\textrm{rk}}
\newcommand{\sch}{\textrm{Sch}}
\newcommand{\EPR}{|\textrm{EPR}\rangle}
\newcommand{\TEPR}{|\Phi^3\rangle}
\newcommand{\GHZ}{|\textrm{GHZ}\rangle}
\newcommand{\defeq}{\stackrel{\textrm{def}}{=}}
\newcommand{\Span}{\mathrm{span}}
\newcommand {\E } {{\mathcal{E}}}
\newcommand{\In}{\mathrm{in}}
\newcommand{\Out}{\mathrm{out}}
\newcommand{\local}{\mathrm{local}}
\newcommand {\F } {{\mathcal{F}}}
\newcommand {\diag } {{\rm diag}}
\renewcommand{\b}{\mathcal{B}}
\newcommand{\h}{\mathcal{H}}
\renewcommand{\Re}{\mathrm{Re}}
\renewcommand{\Im}{\mathrm{Im}}
\newcommand{\Z}{\sigma_z}
\newcommand{\Clocal}{C^{(0)}_{\local}}
\title{Nonlocal Entanglement Transformations Achievable by Separable Operations}
\author{Eric Chitambar}
\email{echitamb@umich.edu}
\affiliation{Physics Department,
University of Michigan, 450 Church Street, Ann Arbor, Michigan
48109-1040, USA\\ and State Key Laboratory of Intelligent Technology
and Systems,Tsinghua National Laboratory for Information Science and
Technology, Department of Computer Science and Technology, Tsinghua
University, Beijing 100084, China}
\author{Runyao Duan}
\email{dry@tsinghua.edu.cn}
\affiliation{State Key Laboratory of
Intelligent Technology and Systems,Tsinghua National Laboratory for
Information Science and Technology, Department of Computer Science
and Technology, Tsinghua University, Beijing 100084, China\\ and
Center of Quantum Computation and Intelligent Systems, Faculty of
Engineering and Information Technology, University of Technology,
Sydney, NSW 2007, Australia}

\date{\today}
\begin{abstract}
For manipulations of multipartite quantum systems, it was well known
that all local operations assisted by classical communication (LOCC)
constitute a proper subset of the class of separable operations.
Recently, Gheorghiu and Griffiths found that LOCC and general
separable operations are equally powerful for transformations
between bipartite \textit{pure} states.  In this letter we extend
this comparison to mixed states and show that in general separable
operations are strictly stronger than LOCC when transforming a mixed
state to a pure entangled state. A remarkable consequence of our
finding is the existence of entanglement monotone which may increase
under separable operations.
\end{abstract}

\maketitle

The ability for quantum systems to exist in non-classical entangled
states presents new possibilities for computational tasks and
information processing \cite{Mermin-2007a}.  Since entanglement is
easily lost through interactions with the environment, any device
that hopes to utilize the properties of entanglement must ultimately
find a way to protect its machinery from excessive environmental
noise.  As the physical separation between different parts of a
quantum system increases, any sort of manipulation on the system as
a whole without introducing environmental interactions becomes even
more challenging, a consideration obviously relevant to any
long-distance quantum communication.

One solution to this problem is to concede the ability of global
operations on the system and require that the only quantum
operations performed are done so locally by different individuals
having access to just a single part of the system.  While still
allowing parties to share classical information with one another,
manipulation protocols of this type are called \textit{LOCC}, an
acronym for local operations and classical communication.  All LOCC
protocols belong to the class of separable operations
\cite{Bennett-1999a} which has a very compact mathematical
characterization.  This fact turns out to be useful since LOCC impossibility results can be obtained by studying the limitations of separable operations, which Rains has done for the task of pure state distillation \cite{Rains-1997a}.  However, it turns out that not all separable operations can be implemented in the form of
an LOCC protocol.  Specifically, as first demonstrated by the
authors of \cite{Bennett-1999a}, there exist sets of states that can
be distinguished by separable operations but not by LOCC.  Beyond
the fact that all LOCC operations are separable, any further
mathematical description of LOCC protocols is lacking. Furthermore,
until now, the focus on tasks demonstrating the difference of
separable operations and LOCC has been limited to state
distinguishability which is unfortunate because it is precisely
these tasks that expose the complex nature of LOCC protocols and
illustrate their operational limitations.

In this Letter, we compare the transforming abilities of separable
operations versus LOCC when acting on general quantum states and
find the latter to be strictly more powerful. Formally, the most
general state of any quantum system can be described by a density
operator belonging to $\mathcal{L}(\h)$, the space of linear
operators acting on some Hilbert space $\h$.  Any change to the
system's state is referred to as a quantum operation and is
mathematically expressed by a superoperator $\mathcal{E}$ that acts
on $\mathcal{L}(\h)$.  It turns out that every physically realizable
superoperator can be represented by a complete set of ``Kraus''
operators $\{E_i\}_{i=1,\cdots,n}$ so that density operators are
mapped according to $\rho\to\mathcal{E}(\rho)=\sum_{i=1}^n E_i \rho
E^\dagger_i$ where $\sum_{i=1}^n E_i^\dagger E_i=I$
\cite{Schumacher-1996a}.
For operations on $N$-partite systems,
if each $E_i$ can be factored in the form
$E_i=\otimes_{k=1}^NA_{ik}$, the physical process described by
$\{E_i\}_{i=1...n}$ is called a \textit{separable operation}.  As
already noted, the Kraus operators representing every LOCC protocol
can be factored as such and thus all LOCC operations are separable;
it is the converse that fails to be true.

Given two bipartite quantum states $\rho$ and $\sigma$, we ask
whether the existence of a separable operation $\mathcal{E}_{Sep}$
such that $\mathcal{E}_{Sep}(\rho)=\sigma$ necessarily implies the
existence of an LOCC operation $\mathcal{E}_{LOCC}$ such that
$\mathcal{E}_{LOCC}(\rho)=\sigma$.  Gheorghiu and Griffiths have
recently shown that indeed, when both $\rho$ and $\sigma$ are pure
states, existence of one implies the existence of the other
\cite{Gheorghiu-2008a}.  Specifically, the authors prove that any pure state
transformation by separable operations obeys the same necessary and
sufficient conditions of an LOCC transformation, namely those given by
Nielsen \cite{Nielsen-1999a} and more generally by Vidal
\cite{Vidal-1999a}  and Jonathan and Plenio \cite{Jonathan-1999a}.  The first step in learning whether such a result holds for all bipartite states is to consider arbitrary
$\rho$ and examine the question of pure state distillation from a
single mixed state.  In other words, given some mixed state $\rho$,
is it possible to transform $\rho$ into a pure state $\sigma$ under a given class of operations?  

It has previously been shown that for mixed states in $2\otimes 2$ systems,
one cannot obtain a pure entangled state via separable operations
with any positive probability of success \cite{Kent-1998a}.  Below,
we prove that the same result holds for $2\otimes 3$ systems when the probability of success is one.  Thus,
for the task of deterministic pure state
distillation on $2\otimes 2$ and $2\otimes 3$ systems, separable
operations and LOCC can be considered as equally powerful. In sharp
contrast to that, we are able to show that for $3\otimes 3$ systems,
mixed states exist in which pure entanglement can be distilled by
separable operations but not by LOCC.  This naturally leads to the
existence of entanglement monotones that increase under separable
operations thus resolving a conjecture intimated in Ref. \cite{Horodecki-2003a}.

The fact that separable transformations on mixed states are more
powerful than LOCC should not be overly surprising. This is because
mixed states can be regarded as an ensemble of pure states such that
mixed state distillation is equivalent to the bulk transformation
between two sets of pure states where, up to complex coefficients,
the target set consists of only one pure state.  We already know
that when the final set contains more than just a single pure state,
there exist source and target sets transformable by separable
operations but not LOCC.  This is indirectly the main result of Ref.
\cite{Bennett-1999a} and more explicitly implied in Ref.
\cite{Duan-2007a}. We state this observation and (thanks to the
findings in Ref. \cite{Bennett-1999a}) its simple proof here for
definitiveness.

\begin{lemma}
\label{lm1} Let $S$ be any set of states distinguishable by
separable operations but not by LOCC. Then there exists sets $S'$
such that $S\to S'$ is achievable by separable operations but not by
LOCC.
\end{lemma}

\textbf{Proof}. Although a non-trivial fact, such sets $S$ do exist
and as pointed out in Ref. \cite{Duan-2007a}, any set of
distinguishable states can always be converted to a set of LOCC
distinguishable states $S'$ through some separable operation. Hence,
conversion $S\to S'$ is impossible through LOCC since by locally
distinguishing the elements of $S'$, the separated parties would
have a way to distinguish the elements of $S$.

When $S'$ contains just a single pure state, this distinguishability
argument no longer applies and we return to the central
investigation of this Letter.  It should be noted that previous research has been done concerning bulk transformations between sets of states using \textit{unrestricted} quantum operations \cite{Chefles-2003a}.  In what follows, we will make
frequent use of a pure state's \textit{Schmidt rank} which is the
minimum number of product states needed to express the given state.
Thus, every entangled state has a Schmidt rank of at least two.  We
begin by examining the differences in separable and LOCC pure state
distillation on lower dimensional systems and find the following:
\begin{theorem}
\label{thm1} Let $\rho$ be any $2\otimes 2$ or $2\otimes 3$ mixed
state. Then for any entangled state $\ket{\phi}$, no separable
operation can transform $\rho$ into $\op{\phi}{\phi}$ with probability one.  Hence, deterministic pure state distillations on $2\otimes 2$ and $2\otimes 3$ systems cannot be achieved by either separable operations or LOCC.
\end{theorem}

\textbf{Proof}.  As Kent has already proven this to be true for
$2\otimes 2$ systems \cite{Kent-1998a}, it is enough to examine
$2\otimes 3$ systems.  Suppose there exists some separable operation
with Kraus operators $\{A_k\otimes B_k\}_{k=1...n}$ that can perform
the indicated transformation.
Without loss of generality, we may assume the rank of $\rho$ is two
and let $\ket{\psi_1}$ and $\ket{\psi_2}$ denote its orthogonal
eigenstates.  Note that the separable operation must also transform any superposition of $\ket{\psi_1}$ and $\ket{\psi_2}$ into $\ket{\phi}$ which implies that $\ket{\psi_1}$ and $\ket{\psi_2}$ each have Schmidt rank at least two.  Then there exists some operator $A\otimes B$ in the set
of Kraus operators such that $(A\otimes B)\ket{\psi_1}=c\ket{\phi}$
and $(A\otimes B)\ket{\psi_2}=d\ket{\phi}$ where $(i)$ both $c$ and
$d$ is nonzero, or $(ii)$ either only one of $c$ or $d$ is nonzero.
Consider case $(i)$.  Since $A$ must be full rank, we have that
$d(I\otimes B)\ket{\psi_1}=c(I\otimes B)\ket{\psi_2}$, or
equivalently $(I\otimes B)(d\ket{\psi_1}-c\ket{\psi_2})=0$. But then
$d\ket{\psi_1}-c\ket{\psi_2}$ is a product vector, which is
impossible since any state in the linear span of $\ket{\psi_1}$ and
$\ket{\psi_2}$ should be entangled. Consider now case $(ii)$ and
without loss of generality, take $c$ to be nonzero. Then $A$ must be
full rank which means for $d$ to be zero, $B$ must be rank one.  But
this is impossible since $\ket{\phi}$ has Schmidt rank two.
\hfill$\blacksquare$

We now move onto the larger dimensional state space of $3\otimes 3$
and show that separable and LOCC pure state distillation abilities
are no longer the same.  The following theorem is the main result of
this Letter.
\begin{theorem}
\label{thm2} (a) Let $\rho$ be any $3\otimes 3$ mixed state and let
$\ket{\phi}$ be any entangled pure state. Then it is impossible to
deterministically convert $\rho$ to $\op{\phi}{\phi}$ by LOCC. (b)
There are $3\otimes 3$ mixed state $\rho$ and pure entangled state
$\ket{\phi}$ such that the transformation of $\rho$ to $\ket{\phi}$
can be achieved with certainty by some separable operation.  Thus,
separable operations are strictly stronger than LOCC for
entanglement transformations between mixed states and pure states.
\end{theorem}

\textbf{Proof}.  $(a)$ Without loss of generality, we may assume
that $\rho$ is of rank two and let $\ket{\psi_1}$ and $\ket{\psi_2}$
be its orthogonal eigenstates. Again note that the stated conversion is
possible if and only if there exists an LOCC protocol $\mathcal{E}$
such that
$\mathcal{E}(\op{\psi_1}{\psi_1})=\mathcal{E}(\op{\psi_2}{\psi_2})=\op{\phi}{\phi}$.
In other words, $\E$ transforms $\ket{\psi_1}$ and $\ket{\psi_2}$
into $\ket{\phi}$ simultaneously. Now, every bipartite LOCC protocol
consists of alternating rounds where one party, say Alice, makes a
measurement on her subsystem and then reports the result to Bob.  In
the next round, Bob selects a particular measurement to make on his
subsystem based on the information given by Alice.  The outcome of
this measurement is reported back to Alice and the cycle repeats.
Hence a tree of all possible outcomes emerges in which each branch
represents a series of operators alternately applied by Alice and
Bob. Specifically, let $N_i$ index all the branches after the
$N^{th}$ round of measurement so that $A^{N_i}\otimes B^{N_i}$
describes the net operations throughout all $N$ rounds along branch
$N_i$.  We first prove the following claim: for any branch $N_i$,
$A^{N_i}\otimes B^{N_i}\ket{\psi_1}\not=0$ if and only if
$A^{N_i}\otimes B^{N_i}\ket{\psi_2}\not=0$.  By induction, assume it
to be true. Then suppose Alice (or Bob) makes the next measurement.
For any $A^j$, let $(N+1)_{ji}$ correspond to the branch
$(A^j\otimes I)(A^{N_i}\otimes B^{N_i})$.  Since this is a
deterministic transformation, $A^{N_i}\otimes
B^{N_i}\ket{\psi_1}\not=0$ implies $A^{N_i}\otimes
B^{N_i}\ket{\psi_1}$ and hence (by assumption) $A^{N_i}\otimes
B^{N_i}\ket{\psi_2}$ must both be states with Schmidt rank at least
two.  Furthermore, if $A^j$ is applied with a nonzero probability on
\textit{either} state, it must also be at least rank two, and thus
$A^j\otimes I$ cannot eliminate any state with Schmidt rank at least
two.  As a result, we have $A^{(N+1)_{ji}}\otimes
B^{(N+1)_{ji}}\ket{\psi_1}\not=0$ if and only if
$A^{(N+1)_{ji}}\otimes B^{(N+1)_{ji}}\ket{\psi_2}\not=0$. Clearly
this argument also applies when $N=0$, so we have proven the claim.

Next consider any branch in the protocol that transforms
$\ket{\psi_1}$ to $\ket{\phi}$.  As just proven, $\ket{\psi_2}$ must
also be transformed to $\ket{\phi}$ along this branch.  Since
$\ket{\psi_1}$ and $\ket{\psi_2}$ are linearly independent but
ultimately become the same state, there must be some round $N$ on
this branch such that $\ket{\psi^{N-1}_1}$ and $\ket{\psi^{N-1}_2}$
are linearly independent, but $A\otimes
I\ket{\psi^{N-1}_1}=c(A\otimes I)\ket{\psi^{N-1}_2}$ where $c\not=0$
is some complex scalar.  Here, $\ket{\psi^{N-1}_{1(2)}}$ denotes the
resultant state of the system originally in state
$\ket{\psi_{1(2)}}$ after round $N-1$, and it is assumed without
loss of generality that Alice makes the $N^{th}$ round measurement.
Because both $\ket{\psi^{N-1}_{1}}$ and $\ket{\psi^{N-1}_{2}}$ can
be converted into $\ket{\phi}$ by the same LOCC protocol, it follows
by linearity of the protocol that any superposition of
$\ket{\psi^{N-1}_{1}}$ and $\ket{\psi^{N-1}_{2}}$ must be entangled.
But then it is impossible that $A\otimes
I\big(\ket{\psi^{N-1}_1}-c\ket{\psi^{N-1}_2}\big)=0$ since $A$ is of
rank at least two, and we arrive at a contradiction.

$(b)$ We give an explicit separable transformation. Consider the
state $\rho=p\op{\psi_1}{\psi_1}+(1-p)\op{\psi_2}{\psi_2}$ where
$\ket{\psi_1}=\frac{1}{\sqrt{2}}(\ket{01}+\ket{10})$ and
$\ket{\psi_2}=\frac{1}{\sqrt{2}}(\ket{02}+\ket{20})$.  Work by Cohen
reveals that these states cannot be distinguished by LOCC while
preserving their Schmidt rank, but at the same time, such a
discrimination is possible by separable operations
\cite{Cohen-2007a}.  By modifying a POVM given in Ref.
\cite{Cohen-2007a}, we obtain a separable operation $\E$ described
by the following set of Kraus operators $\{E_k:k=1,\cdots,6$\}:
\begin{eqnarray}
E_1&=&\sqrt{\alpha}(\sqrt{\beta}\op{0}{0}+\op{1}{1})\otimes(\sqrt{\beta}\op{0}{1}+\op{1}{0}),\nonumber\\
E_2&=&\sqrt{\alpha}(\op{1}{0}+\sqrt{\beta}\op{0}{1})\otimes(\op{1}{1}+\sqrt{\beta}\op{0}{0}),\nonumber\\
E_3&=&\sqrt{\alpha}(\sqrt{\beta}\op{0}{0}+\op{1}{2})\otimes(\sqrt{\beta}\op{0}{2}+\op{1}{0}),\nonumber\\
E_4&=&\sqrt{\alpha}(\op{1}{0}+\sqrt{\beta}\op{0}{2})\otimes(\op{1}{2}+\sqrt{\beta}\op{0}{0}),\nonumber\\
E_5&=&\op{1}{1}\otimes(\sqrt{2\alpha\beta}\op{1}{1}+\op{2}{2}),\nonumber\\
E_6&=&\op{2}{2}\otimes(\op{1}{1}+\sqrt{2\alpha\beta}\op{2}{2}),
\end{eqnarray}
where $\alpha=\frac{2-\sqrt{3}}{4}$ and $\beta=2+\sqrt{3}$.  One can
readily verify that $\E$  deterministically transforms $\rho$ into
$\op{\phi}{\phi}$ with
$\ket{\phi}=\sqrt{\alpha}(\beta\ket{00}+\ket{11})$.\hfill$\blacksquare$

Theorem \ref{thm2} leads directly to the existence of entanglement
monotones that increase under separable operations.  Recall that an
entanglement monotone is some function mapping density operators to
the real numbers that does not increase on average under LOCC
transformations \cite{Vidal-2000a}.  As demonstrated in Ref.
\cite{Vidal-2000a}, any deterministic transformation not achievable
by LOCC necessarily implies the increase in some entanglement
monotone.  Thus, we obtain the following consequence.
\begin{corollary}
\label{cor1} There exist entanglement monotones that can increase
under separable operations.
\end{corollary}

The results of Ref. \cite{Gheorghiu-2008a} show that any
entanglement monotone defined by a convex roof extension of some
pure state measure will necessarily be monotonic under separable
operations. Consequently, the monotone that increases under
separable operations must be defined in another manner such as the
one provided in Theorem 1 of Ref. \cite{Vidal-2000a}.  Specifically,
for any states $\sigma$ and $\rho$, let $P^{max}_\sigma(\rho)$ be
the maximum probability of obtaining $\sigma$ from $\rho$ by LOCC.
Then $P^{max}_\sigma(\rho)=\underset{\mathcal{E}}{\text{max}} \sum_{i=1}^n
p_i P^{max}_\sigma(\rho_i)$ where the maximum is taken over all LOCC
protocols $\mathcal{E}$ transforming $\rho$ into an ensemble
$\{(p_i, \rho_i):i=1,\cdots, n\}$ where output state $\rho_i$ occurs
with probability $p_i$. As a result, $P^{max}_\sigma$ cannot
increase on average under LOCC since the maximum average after any
LOCC protocol is precisely $P^{max}_\sigma$.  The source and target
states of Theorem \ref{thm2} provide an example of when
$P^{max}_\sigma$ increases under separable operations.

No matter how we define our entanglement monotone, product states
will always have the lowest possible amount of entanglement since
any product state can be obtained by LOCC from any original state.
Because separable operations map product states to product states,
they will never be able to increase the entanglement of systems
originally unentangled, regardless of what monotone we use.  Thus it
is interesting to note that only if the original state is entangled
can separable operations increase its entanglement according to some
entanglement monotone, and as now evident, in some cases it does.

Many open questions exist concerning the items discussed in this
Letter. It would be desirable to know if separable and LOCC
transformations are equally strong on $2\otimes n$ systems.  For
$n>3$, it is not difficult to construct mixed states from which pure
entanglement can be deterministically distilled.  However in all
such examples we found an LOCC protocol that could achieve the same
transformation.  Another relevant question is whether separable
operations are more powerful than LOCC on $2\otimes 2$ and $2\otimes
3$ systems when the target state is mixed.  With the class of
separable operations being so large, this seems highly plausible.

When viewed in conjunction with Cohen's work \cite{Cohen-2007a}, our
results suggest that a critical difference between LOCC and
separable operations is their ability to preserve entanglement when
acting on multiple pure states.  Specifically, since LOCC protocols
consist of multi-round measurements, for the tasks we consider the Schmidt rank of a state
must be preserved after each stage, whereas with separable
operations, it only needs to be preserved after a ``single shot''.
Thus it is no surprise that the ability to perform certain quantum
operations by LOCC is strictly less than separable operations.  On
the other hand, the phenomenon of nonlocality without entanglement
\cite{Bennett-1999a}, i.e., the existence of orthogonal product
states indistinguishable by LOCC but not by separable operations,
implies that preservation of entanglement cannot be the whole story
to the difference between LOCC and separable operations. As further
effort is put forth to understand the complicated nature of LOCC
protocols, we hope that the results presented here generate useful
tools and considerations to assist in this endeavor.  At the very
least, we have introduced a new practical scenario distinct from
state distinguishability in which the capabilities of LOCC protocols
cannot simply be obtained by studying separable operations.

We thank Yaoyun Shi and Andreas Winter for helpful discussions and
comments.  This work was partially supported by the National Natural
Science Foundation of China (Grant Nos. 60702080, 60736011, and
60621062), the FANEDD under Grant No. 200755, and the Hi-Tech
Research and Development Program of China (863 project) (Grant No.
2006AA01Z102). Eric Chitambar was also partially supported by the
National Science Foundation of the United States under
Awards~0347078 and 0622033.

\bibliography{QuantumBib}

\end{document}